\documentclass{elsart}

\newcounter{bla}

\usepackage{graphics}
\usepackage{graphicx}

\begin{document}
\begin{frontmatter}

\title{BOKASUN: a fast and precise numerical program to calculate the 
 Master Integrals of the two-loop sunrise diagrams.\thanksref{paper}}

\author[a]{Michele Caffo},
\author[b]{Henryk Czy\.z\thanksref{author}},
\author[b]{Micha{\l} Gunia},
\author[a]{Ettore Remiddi}

\thanks[paper]{Work supported in part by 
EU 6th Framework Programs under contracts MRTN-CT-2006-035482 
(FLAVIAnet) and MRTN-CT-2006-035505 (HEPTOOLS). }
\thanks[author]{Corresponding author}

\address[a]{INFN and Dipartimento di Fisica dell'Universit\`a, Bologna, Italy}
\address[b]{Institute of Physics, University of Silesia, Katowice, Poland}

\begin{abstract}
   We present the program BOKASUN for fast and precise evaluation
 of the Master Integrals of the two-loop self-mass sunrise diagram
 for arbitrary values of the internal masses and the external four-momentum.
 We use a combination of two methods: a Bernoulli 
 accelerated series expansion 
 and a Runge-Kutta numerical solution of a system of linear differential
 equations. 
\begin{flushleft}

\end{flushleft}


\end{abstract}

\end{frontmatter}
\newcommand{\F}[1]{F_#1(n,m_1^2,m_2^2,m_3^2,p^2)} 
\newcommand{\dnk}[1]{ d^nk_{#1} } 
\newcommand{\Fn}[2]{F_{#1}^{(#2)}(m_1^2,m_2^2,m_3^2,p^2)}
\newcommand{\D}{D(m_1^2,m_2^2,m_3^2,p^2)} 


{\bf PROGRAM SUMMARY}

\begin{small}
\noindent
{\em Program Title: }    BOKASUN                                      \\
{\em Journal Reference:}                                      \\
{\em Catalogue identifier:}                                   \\
{\em Licensing provisions: }    none                               \\
{\em Programming language: }    FORTRAN77                               \\
{\em Computer: } any computer with FORTRAN compiler accepting FORTRAN77 standard;
 tested on various PC's with LINUX                                              \\
{\em Operating system: }  LINUX                                     \\
{\em RAM:} 120 kbytes                                              \\
{\em Classification: 4.4}                                         \\

{\em Nature of problem:}\\
   Any integral arising in the evaluation of the two-loop sunrise Feynman 
  diagram can be expressed in terms of a given set of Master Integrals, 
  which should be calculated numerically.
  The program provides with a fast and precise evaluation method of the 
  Master Integrals for arbitrary (but not vanishing) masses and 
  arbitrary value of the external momentum.
   \\ \\
{\em Solution method:}\\
  The integrals depend on three internal masses and the external momentum 
  squared $p^2$. The method is a combination of an accelerated expansion 
  in $1/p^2$ in its (pretty large!) region of fast convergence and of a 
  Runge-Kutta numerical solution of a system of linear differential 
  equations.  
   \\ \\
{\em Running time:}\\ To obtain 4 Master Integrals on PC with 2 GHz processor
 it takes 
 3 $\mu$s for series expansion with 
 calculated in advance coefficients, 80 $\mu$s for series expansion without 
 calculated in advance coefficients, from few seconds up to few minutes 
 for Runge-Kutta method
 (depending on the required accuracy and the values of the physical 
 parameters).\\
   \\

\end{small}


\hspace{1pc}
{\bf LONG WRITE-UP}

\section{Introduction}

 The sunrise diagram with arbitrary masses is one of the basic ingredients 
 of any two-loop
 calculation, and its fast numerical evaluation is of direct interest 
 in Monte Carlo simulation programs.
 Many procedures for a precise evaluation of the Master Integrals (MI's) 
 can be found in the literature 
 \cite{Berends:1993ee,Berends:1994ed,Bauberger:1994by,Ghinculov:1994sd,
Post:1996gg,Groote:1999cx,Amoros:1999dp,Passarino:2001wv,Passarino:2001jd,Laporta:2001dd,Caffo:2002ch,Martin:2003qz,Caffo:2003ma,Martin:2005qm}.
 In \cite{Pozzorini:2005ff} a fast and precise series expansion was proposed 
 in the much simpler case of equal internal masses. 
 In this paper we adapt the method to the arbitrary mass case by considering 
 the accelerated expansion in inverse powers of $p^2$, which provides with 
 a fast and precise convergence in a wide region covering the biggest part of 
 $p^2$ values. 
 The expansion does not work in a relatively small region (roughly from 
 $p^2= 0$ to the physical threshold). 
 A similar expansion in $p^2$ around the regular point $p^2= 0$ (the first 
 terms where given in \cite{Caffo:1998du}) is unpractical due to the severe 
 numerical instability of the coefficients of the expansion \cite{HC}, 
associated with the presence of nearby pseudothresholds 
 (this feature is peculiar to the arbitrary mass case, as opposed to the 
 equal mass limit).   
 In the regions were the expansion in $1/p^2$ does not work we use the 
 Runge-Kutta algorithm developed in 
\cite{Caffo:2002ch,Caffo:2002we,Caffo:2002wm}
 to obtain the Master Integrals (MI's) of the sunrise diagram 
as the numerical solutions of a suitable system of linear 
differential equations. 
 The execution time is of the order of a few seconds (minutes)
  when the Runge-Kutta 
 method is used (depending  on the values of $p^2$ and the masses
 and also on the required precision), 
 while it drops to about $80\  \mu$s when the new expansion applies
 and to $3\  \mu$s if the expansion coefficients are calculated in advance 
 (all CPU times are given for a 2 GHz PC).  
 
\section{The notation}

We use here the same notation and definitions as in \cite{Caffo:2002ch}, which
 we recall shortly for convenience of the reader.
 The four Master Integrals (MI's) related to the 
general massive 2-loop sunrise self-mass diagram in $n$ continuous 
dimensions and with fully Euclidean variables are defined as

\begin{eqnarray} 
 \F{j}
      &=& \frac{ \mu^{8-2n}}{((2\pi)^{n-2})^2 }\nonumber \\  
      && {\kern-120pt} \int \dnk{1} \int \dnk{2} \; 
      \frac{ 1 } 
           { (k_1^2+m_1^2)^{\alpha_1(j)} (k_2^2+m_2^2)^{\alpha_2(j)} 
             ( (p-k_1-k_2)^2+m_3^2 )^{\alpha_3(j)} } \ , \nonumber \\ 
\label{1} \end{eqnarray} 

where $j=0,1,2,3$ refers to the 4 MI's; for $ j=0 $, $ \alpha_i(j=0)=1 $ 
for $ i=1,2,3 $; for $ j>0 $, $ \alpha_i(j)=2 $ when $i=j $ and 
$ \alpha_i(j) = 1 $ when $ i \ne j $. 

The mass scale is chosen as 
\begin{equation} 
 \mu = m_1+m_2+m_3 \ , 
\label{mu} \end{equation} 
which comes out to be the appropriate 
mass scale parameter for the numerical discussion. 
The expansion of the MI's around \( n = 4 \) has the form \cite{Caffo:1998du}
\begin{eqnarray} 
 && {\kern-30pt}\F{j} = C^2(n) \Biggl\{  \frac{1}{(n-4)^2} \Fn{j}{-2} 
            \nonumber \\ 
 &&  + \frac{1}{(n-4)}   \Fn{j}{-1} 
                  + \Fn{j}{0} + {\cal O} (n-4) \Biggr\} \ . \nonumber \\ 
\label{39a} \end{eqnarray} 
where the coefficient $C(n)$  
\begin{equation} 
C(n) = \left(2 \sqrt{\pi} \right)^{(4-n)} \Gamma\left(3-\frac{n}{2}\right) 
                                                               \ , 
\label{16a} \end{equation} 
not to be expanded, can be replaced by its value $C(4)=1$, at $n = 4$, 
when multiplying a function regular in \( (n-4) \). 
The coefficients of the poles in \( (n-4) \) of \( \F{j} \) are 
known in closed analytic form \cite{Caffo:1998du,Caffo:2002ch}, 
and are not reconsidered here, as from now on we deal only with 
the finite parts $\Fn{j}{0}$ of the MI's.

It is convenient to use reduced masses and reduced external invariant 
\begin{equation} 
m_{i,r} \equiv \frac{m_i}{m_1+m_2+m_3} \quad, \quad 
p^2_r \equiv \frac{p^2}{(m_1+m_2+m_3)^2} \ , 
\label{red1} \end{equation} 
together with a dimensionless version of \( \F{0} \), defined by 
\begin{equation} 
 \F{{0,r}} \equiv \frac{\F{0}}{(m_1+m_2+m_3)^2} \ ; 
\label{red2} \end{equation} 
as the other Master Integrals are already dimensionless, the values 
of all the functions are also dimensionless. 
In terms of the new variables \(p^2_r, m_{i,r}\) the threshold is located at  
\( p_{th,r}^2  = -1 \). 

Moreover we do not write anymore, for short, the arguments of the functions
and the superscript $(0)$, so we set 
\( F_0 \equiv \Fn{{0,r}}{0} \) and \( F_j \equiv \Fn{j}{0} ,\ j=1,2,3 \).

\section{Handling of the asymptotic expansion}

The asymptotic expansion at large $p_r^2$ values was proposed in
\cite{Caffo:1998du}, but only few terms were given there explicitly.
It can be written symbolically as

\begin{eqnarray}
  F_i = \Sigma_{i,0} + \log(p_r^2) (\Sigma_{i,1}) 
                     + \log^2(p_r^2) (\Sigma_{i,2}) \ , 
 \ \ \ i=0,1,2,3 \ ,
 \label{ff}
\end{eqnarray}

where $\Sigma_{i,j} $ are power series in $1/p_r^2$.
 For the purpose of the program presented here, up to 18 terms in inverse
 powers of $p_r^2$ were evaluated. That required the solution of a system 
 of four linear equations per set of coefficients.
  Each coefficient in the expansion
 is a polynomial in three masses and logarithms of the masses.
 As the length of the coefficients is growing with the growing inverse power
 of $p_r^2$ we have made an effort to shorten the expressions using symmetric
 polynomials for $F_0$. 
 The polynomials are calculated once and used several
 times in the evaluation of the coefficients. For the $F_i$ 
 ($i=1,2,3$) we found out that
 the most effective way to simplify the coefficients is to make use of the
 relation $\F{i}= -\partial \F{0} / \partial m_i^2$ 
 after the simplification of the
 coefficients of $F_0$. All that resulted not only in shortening of the
 expressions, but what is more important, in significant (about 10 times)
 gain in CPU time necessary for the calculation of the sunrise master 
integrals.

 To speed up the convergence of the series and to enlarge the convergence 
region we use the Bernoulli change of variables, introduced in 
 \cite{'t Hooft:1978xw} and systematically used in 
 \cite{Gehrmann:2001jv,Pozzorini:2005ff}, 
 separately for each of the series
  $\Sigma_{i,j}$ from Eq.(\ref{ff}).
   Each of the  series $\Sigma$ (we drop here the subscript
  to shorten the expressions)
\begin{eqnarray}
  \Sigma = \sum^{\infty}_{n=0} a_n \left(\frac{1}{p_r^2}\right)^n ,
\end{eqnarray}
after the (Bernoulli) change of variable
\begin{eqnarray}
y=\texttt{log}\left(1+\left(\frac{1}{p_r^2}\right)\right) 
\end{eqnarray}
becomes
\begin{eqnarray}
 \Sigma =a_0 + \sum^\infty _{l=0}\frac{1}{l!}y^lb_l \ ,
\end{eqnarray}
with
\begin{eqnarray}
b_l = \sum^{l}_{n=1} a_n (-1)^{n}\sum^{n}_{k=0}\frac{n!}{(n-k)!k!}(-1)^k k^l \ .
\label{coefbern}
\end{eqnarray}

We obtain an expansion in $y$ which, with 18 terms, provides with double
precision (real*8)
results for all values of $p_r^2$ outside the interval
[~-1.5,0.5] and for arbitrary masses. The precision of the result
 is estimated by taking the ratio of the last term to the sum of all the 
terms. As a matter of fact the double precision accuracy is obtained
 for particular values of masses also in a wider region of $p_r^2$ 
 (see next section). The estimation of the error was checked against
 the results of a Runge-Kutta method of comparable precision.
It is to be recalled here that the convergence of the expansion
 in powers of $1/p_r^2$ is superior to the Runge-Kutta approach for
 large values of $p_r^2$, so that the check is a really stringent
 one in the region $-1.5 < p_r^2 < -1$ and $0 < p_r^2 < 0.5$ at the
 borderline of the convergence of the expansion.

As the coefficients of the Bernoulli accelerated series are obtained
 numerically, we have checked the numerical stability of the procedure.
 The expected cancellations occurring in Eq.(\ref{coefbern}) were never
 affecting the accuracy of the result and the formulae remained numerically
 stable at variance with the formulae for the expansion at $p^2=0$,
 which were thus not used in the program.

For the values of $ p^2_r $ for which the asymptotic expansion does not work  
 we use 
 the direct numerical solution of the system of differential equations
 by means of the Runge-Kutta method described in details in \cite{Caffo:2002ch}.
The algorithm
 works relatively fast
 in the region of small  $p_r^2$ values.
  Thus the combination of the methods, 
used in the present program, allows for the fast and accurate evaluation
 of the sunrise Master Integrals for all values of $p_r^2$.

\section{The outline of the program}

\begin{figure}[h]
\begin{center}
\includegraphics[scale=0.8]{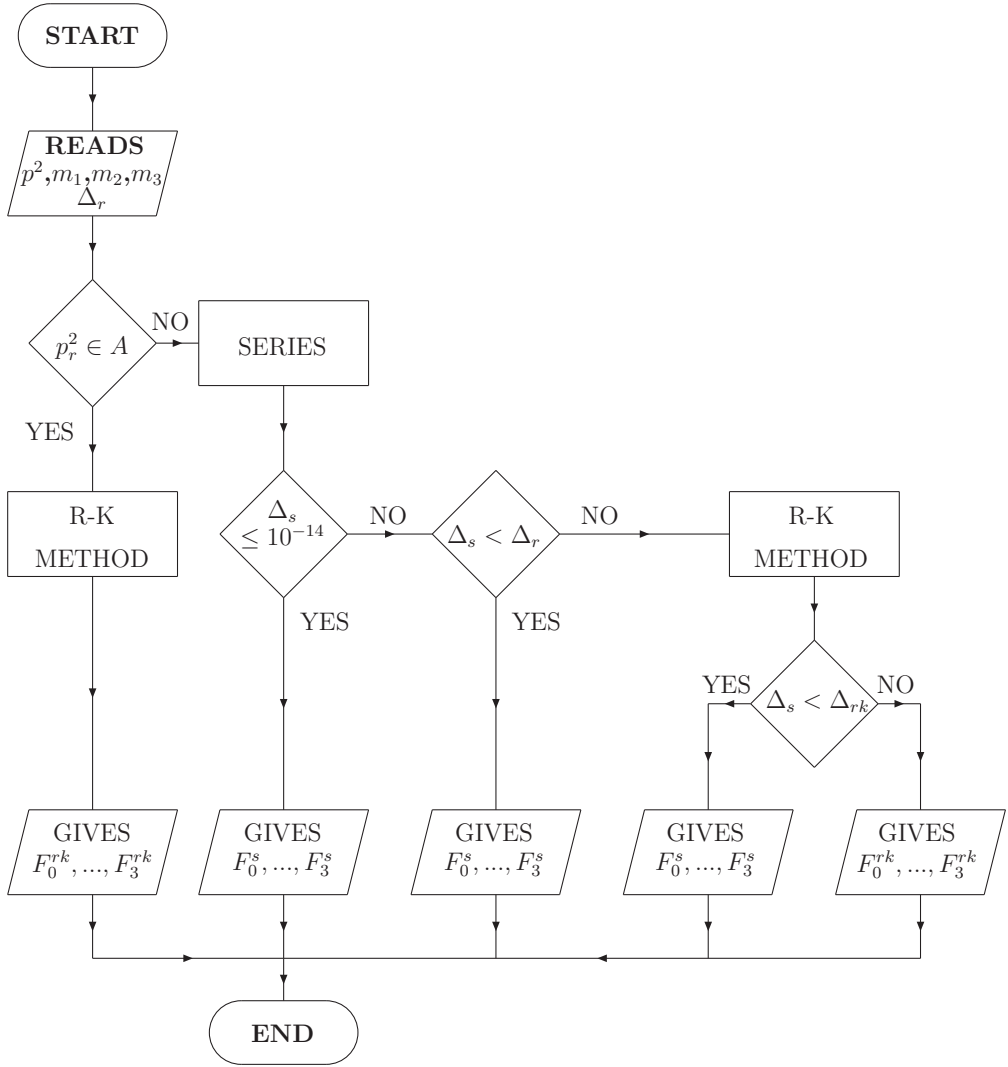}
\end{center}
\caption{The flowchart  of the BOKASUN program (see the text for details).
  $\Delta_s$ ($\Delta_{rk}$): accuracy obtained by the expansion 
 (Runge-Kutta)
 method. $F_n^s$ ($F_n^{rk}$), $n=0,...,3$:  results obtained by means of
 the expansion (Runge-Kutta) method. The $\Delta_s < \Delta_{rk}$ condition
 is checked separately for the real and  the imaginary part of each MI. Thus,
 when the MI's are calculated by both methods,  the result with better
 accuracy is given in the output.  }
\label{program}
\end{figure}

The way the program works is shown  schematically in Fig. \ref{program}.
First, the program reads from the file  {\tt input\_BOKASUN.dat}
the  values of $p^2, m_1,m_2,m_3$,
the required relative accuracies (denoted as 
 $\Delta_r$, $\Delta_r=({\tt acc(1),acc(2),acc(3)})$ ) 
 for the real parts, the imaginary parts and
 the moduli of the four functions, which the user wishes to calculate.
 The program checks if the value of the squared rescaled four momentum ($p_r^2$)
 is within the interval $A=[-1,0]$, where the series expansion is not valid.
 For $p_r^2\in A$ the program uses the Runge-Kutta method, described in 
  \cite{Caffo:2002ch}, and gives the values of all functions 
 $ F_n$.
 The program tries to reach the accuracy asked by the user and gives the
 values with the best obtained accuracy, even if the required accuracy
 was not reached. Outside the interval $A$ the program evaluates first the
 value of the functions $ F_n$
  by using the asymptotic expansion described in the previous section.
  If the estimated relative accuracies for real and imaginary parts 
  are lower than $10^{-14}$ or
  the accuracies are better (or equal) than (to) the accuracies
  asked by the user
  the program writes the $ F_n$
  with their relative accuracies for real and
  imaginary parts. If any of the accuracy requirements is not met
  the program uses Runge-Kutta method to calculate
 all functions $F_n,\ n=0,1,2,3$. The result with better accuracy is
 written as the output.

There are two subroutines to calculate the MI's:

{\tt bokasun}, where the series expansion coefficients are calculated for
 each call, and

{\tt bokasun\_s}, where the series expansion coefficients are calculated
 in advance in the subroutine {\tt prepare\_store}

The second option is useful when one needs to calculate the MI's for fixed
 masses and various $p^2$ values. It speeds up the calculations about 25 times.

 Both subroutines are called with parameters:

 {\tt p2,m1in,m2in,m3in,acc,Fn,deltare,deltaim}

declared as

{\tt
      real*8 p2,m1in,m2in,m3in,acc(3) \\
      complex*16 Fn(0:3) \\
      real*8 deltare(0:3),deltaim(0:3)\\}

{\tt p2} is the square of the four momentum

 {\tt m1in,m2in,m3in} are the internal masses

 {\tt acc(1)} is the required relative precision for the real parts of 
the MI's 

 {\tt acc(2)} is the required relative precision for the imaginary parts of 
the MI's

 {\tt acc(3)} is the required relative precision for the modulus of the MI's;
 used only for Runge-Kutta method

 {\tt Fn(i), i=0,1,2,3} finite part of the ith MI

 {\tt deltare(i) ( deltaim(i))} relative accuracy of the real (imaginary)
 part of the ith MI

\section{Tests of the program and typical run times}

 In \cite{Caffo:2002ch} many tests  of the Runge-Kutta method were performed
 in all regions of the $p^2$ values. Comparisons
 have shown an excellent agreement
 between the code developed in \cite{Caffo:2002ch} and the values 
 published in \cite{Berends:1993ee,Passarino:2001wv}.
 In \cite{Martin:2003qz} the author states also the complete agreement
 with \cite{Caffo:2002ch} of his code, published later as a part of 
  \cite{Martin:2005qm}.  
 In view of these comparisons we have just checked that the new part
 of the program,
 which uses the large $p_r^2$ expansion gives results 
 which are in agreement with the Runge-Kutta method.
 As matter of fact, when the expansion in $ 1/p^2_r $ applies, 
our program reaches a precision of $ 10^{-14} $ or better, which is 
higher than the other available programs, so that a direct 
comparison up to that precision was in general not possible. 
Nevertheless, extensive comparisons were made, limited to the relative 
precision of  $10^{-11}$ (or slightly better; that is 
 the maximum precision which one can reach with the Runge-Kutta
  method for large $p_r^2$ values) for various sets of masses. 
 The results were always in agreement within the errors.
 Machine precision comparisons were possible with \cite{Pozzorini:2005ff}
 in the equal mass case, and an excellent agreement was found. 

 The main gain in using the series 
 expansion, whenever possible, is the reduction of 
 CPU time necessary to obtain the result.
 The CPU time (on a laptop with Intel Centrino Duo T7400 2.16 
 GHz processor) necessary for calculation of the Master Integrals 
 with the double precision machine accuracy, with the expansion method,
 is about $8\cdot 10^{-5}\  s$, which reduces to 
 about $3\cdot 10^{-6}\  s$ when the expansion coefficients
 are calculated in advance. With the Runge-Kutta method it takes
 $10\ s$ at $p_r^2=0.1$ and $1800 \ s$ at $p_r^2=10$ to
 obtain the relative accuracy of $10^{-11}$.


\hspace{1pc}
{\bf TEST RUN OUTPUT}

\bigskip
The distributed version of the program contains also a code of the test
 run, which uses both subroutines {\tt bokasun} and  {\tt bokasun\_s}.
 It provides also with an example of using the BOKASUN program.
 It reads the input parameters for the test run from the file
{\tt input\_BOKASUN.dat } and appends the results of the MI's with the
 obtained relative accuracies to files {\tt f0.dat} ($F_0$),
 {\tt f1.dat} ($F_1$), {\tt f2.dat} ($F_2$)
 and  {\tt f3.dat} ($F_3$). In the files {\tt f0.dat.ref},
{\tt f1.dat.ref}, {\tt f2.dat.ref} and {\tt f3.dat.ref},
 distributed together with the program, the results expected for the test run
 are given. The test run takes about 6 minutes CPU as it calculates many points
 using the Runge-Kutta method.

{\bf Acknowledgments}

Henryk Czy\.z is
  grateful for the support and the kind hospitality
of the INFN and Dipartimento di Fisica dell'Universit\`a di Bologna.


\begin{thebibliography}{99}

\bibitem{Berends:1993ee}
  F.~A.~Berends, M.~Buza, M.~Bohm and R.~Scharf,
  Z.\ Phys.\  C {\bf 63} (1994) 227.

\bibitem{Berends:1994ed}
  F.~A.~Berends and J.~B.~Tausk,
  Nucl.\ Phys.\  B {\bf 421} (1994) 456.

\bibitem{Bauberger:1994by}
  S.~Bauberger, F.~A.~Berends, M.~Bohm and M.~Buza,
  Nucl.\ Phys.\  B {\bf 434} (1995) 383
  [arXiv:hep-ph/9409388].

\bibitem{Ghinculov:1994sd}
  A.~Ghinculov and J.~J.~van der Bij,
  Nucl.\ Phys.\  B {\bf 436} (1995) 30
  [arXiv:hep-ph/9405418].


\bibitem{Post:1996gg}
  P.~Post and J.~B.~Tausk,
  Mod.\ Phys.\ Lett.\  A {\bf 11} (1996) 2115
  [arXiv:hep-ph/9604270].

\bibitem{Groote:1999cx}
  S.~Groote, J.~G.~K\"orner and A.~A.~Pivovarov,
  Eur.\ Phys.\ J.\  C {\bf 11} (1999) 279
  [arXiv:hep-ph/9903412].
  Nucl.\ Phys.\  B {\bf 542} (1999) 515
  [arXiv:hep-ph/9806402].

\bibitem{Amoros:1999dp}
  G.~Amoros, J.~Bijnens and P.~Talavera,
  Nucl.\ Phys.\  B {\bf 568} (2000) 319
  [arXiv:hep-ph/9907264].

\bibitem{Passarino:2001wv}
  G.~Passarino,
  Nucl.\ Phys.\  B {\bf 619} (2001) 257
  [arXiv:hep-ph/0108252].

\bibitem{Passarino:2001jd}
  G.~Passarino and S.~Uccirati,
  Nucl.\ Phys.\  B {\bf 629} (2002) 97
  [arXiv:hep-ph/0112004].

\bibitem{Laporta:2001dd}
  S.~Laporta,
  Int.\ J.\ Mod.\ Phys.\  A {\bf 15} (2000) 5087
  [arXiv:hep-ph/0102033].
  Phys.\ Lett.\  B {\bf 504} (2001) 188
  [arXiv:hep-ph/0102032].


\bibitem{Caffo:2002ch}
  M.~Caffo, H.~Czy{\. z} and E.~Remiddi,
  Nucl.\ Phys.\  B {\bf 634} (2002) 309
  [arXiv:hep-ph/0203256].

\bibitem{Martin:2003qz}
  S.~P.~Martin,
  Phys.\ Rev.\  D {\bf 68} (2003) 075002
  [arXiv:hep-ph/0307101].

\bibitem{Caffo:2003ma}
  M.~Caffo, H.~Czy\.z, A.~Grzeli\'nska and E.~Remiddi,
  Nucl.\ Phys.\  B {\bf 681} (2004) 230
  [arXiv:hep-ph/0312189].

\bibitem{Martin:2005qm}
  S.~P.~Martin and D.~G.~Robertson,
  Comput.\ Phys.\ Commun.\  {\bf 174} (2006) 133
  [arXiv:hep-ph/0501132].

 
\bibitem{Pozzorini:2005ff}
  S.~Pozzorini and E.~Remiddi,
  Comput.\ Phys.\ Commun.\  {\bf 175} (2006) 381
  [arXiv:hep-ph/0505041].

\bibitem{Caffo:1998du}
  M.~Caffo, H.~Czy{\. z}, S.~Laporta and E.~Remiddi,
  Nuovo Cim.\  A {\bf 111} (1998) 365
  [arXiv:hep-th/9805118].

\bibitem{HC} H.~Czy{\. z}, unpublished. 

\bibitem{Caffo:2002we}
  M.~Caffo, H.~Czy{\. z} and E.~Remiddi,
  Nucl.\ Instrum.\ Meth.\  A {\bf 502} (2003) 613
  [arXiv:hep-ph/0211171].


\bibitem{Caffo:2002wm}
  M.~Caffo, H.~Czy{\. z} and E.~Remiddi,
  Nucl.\ Phys.\ Proc.\ Suppl.\  {\bf 116} (2003) 422
  [arXiv:hep-ph/0211178].




 \bibitem{'t Hooft:1978xw}
  G.~'t Hooft and M.~J.~G.~Veltman,
  Nucl.\ Phys.\  B {\bf 153} (1979) 365.



\bibitem{Gehrmann:2001jv}
  T.~Gehrmann and E.~Remiddi,
  Comput.\ Phys.\ Commun.\  {\bf 144} (2002) 200
  [arXiv:hep-ph/0111255].



 \end{thebibliography}
\end{document}